\documentclass[twocolumn, switch]{article} 

\usepackage{preprint}

\usepackage{amsmath, amsthm, amssymb, amsfonts}

\usepackage[numbers,square]{natbib}
\usepackage[utf8]{inputenc}	
\usepackage[T1]{fontenc}	
\usepackage{xcolor}		
\usepackage[colorlinks = true,
            linkcolor = purple,
            urlcolor  = blue,
            citecolor = cyan,
            anchorcolor = black]{hyperref}	
\usepackage{booktabs} 		
\usepackage{nicefrac}		
\usepackage{microtype}		
\usepackage{lineno}		
\usepackage{float}			

\usepackage{lipsum}		

\usepackage{newfloat}
\DeclareFloatingEnvironment[name={Supplementary Figure}]{suppfigure}
\usepackage{sidecap}
\sidecaptionvpos{figure}{c}

\usepackage{titlesec}
\titlespacing\section{0pt}{12pt plus 3pt minus 3pt}{1pt plus 1pt minus 1pt}
\titlespacing\subsection{0pt}{10pt plus 3pt minus 3pt}{1pt plus 1pt minus 1pt}
\titlespacing\subsubsection{0pt}{8pt plus 3pt minus 3pt}{1pt plus 1pt minus 1pt}

\title{Active Matter Invasion}

\usepackage{xwatermark}
\newwatermark[firstpage,color=gray!60,angle=90,scale=0.32, xpos=3.9in,ypos=0]{\href{https://doi.org/10.1039/c9sm01210a}{\color{gray}{article doi: https://doi.org/10.1039/c9sm01210a}}}
\newwatermark[firstpage,color=gray!90,angle=0,scale=0.28, xpos=0in,ypos=-5in]{*correspondence: \texttt{amin.doostmohammadi@physics.ox.ac.uk}}

\usepackage{authblk}

\author[1]{Felix Kempf}
\author[2]{Romain Mueller}
\author[1]{Erwin Frey}
\author[2]{Julia M. Yeomans}
\author[2\thanks{\tt{amin.doostmohammadi@physics.ox.ac.uk}}]{Amin Doostmohammadi}

\affil[1]{Arnold Sommerfeld Center for Theoretical Physics and Center for NanoScience, Department of Physics, Ludwig-Maximilians-Universit\"at M\"unchen - Theresienstr.  37, D-80333 Munich, Germany.}
\affil[2]{The Rudolf Peierls Centre for Theoretical Physics - Clarendon Laboratory, Parks Road, Oxford, OX1 3PU, UK.}

\newcommand{\hdiff}{\ensuremath{\left(h_{\text{max}}-h_{\text{min}}\right)/d}}
\newcommand{\hdiffinva}{\ensuremath{\left(h_{\text{inv}}-h_{\text{max}}\right)/d}}
\renewcommand{\vec}{\mathbf}
\newcommand{\binary}{D}
\newcommand{\coupling}{C_{LQ}}
\newcommand{\Kphi}{K_{\phi}}
\newcommand{\Knem}{K_Q}
\newcommand{\tumbling}{\xi}
\newcommand{\activescale}{\Lambda_{\zeta}}
\newcommand{\Invasion}{\Phi_T}

\newcommand{\ESImovieI}{flows0020.mp4}
\newcommand{\ESImovieII}{flows0030.mp4}
\newcommand{\ESImovieIII}{flows0035.mp4}
\newcommand{\ESImovieIV}{flows0046.mp4}
\newcommand{\ESImovieV}{flows0052.mp4}
\newcommand{\ESImovieVI}{flows0060.mp4}

\newcommand{\ESIsectionreservoir}{A}
\newcommand{\ESIsectionwidth}{B}
\newcommand{\ESIsectionmovies}{C}

\newcommand{\ESIfigwidth}{2}
\newcommand{\ESIfigdefects}{3}

\newcommand{\maintextfigregimes}{2}
\newcommand{\maintextfigflows}{3}
\newcommand{\maintextfigdynamics}{4}
\newcommand{\maintextsectionresultsquantitativeregimeII}{3.2.1~}
\newcommand{\maintextsectionresultsquantitativeregimeIII}{3.2.2~}

\begin{document}

\twocolumn[ 
  \begin{@twocolumnfalse} 
  
\maketitle

\begin{abstract}
Biological active materials such as bacterial biofilms and eukaryotic cells thrive in confined micro-spaces. Here, we show through numerical simulations that confinement can serve as a mechanical guidance to achieve distinct modes of collective invasion when combined with growth dynamics and the intrinsic activity of biological materials. We assess the dynamics of the growing interface and classify these collective modes of invasion based on the activity of the constituent particles of the growing matter. While at small and moderate activities the active material grows as a coherent unit, we find that blobs of active material collectively detach from the cohort above a well-defined activity threshold. We further characterise the mechanical mechanisms underlying the crossovers between different modes of invasion and quantify their impact on the overall invasion speed.
\end{abstract}
\vspace{0.35cm}

  \end{@twocolumnfalse} 
] 



\section{Introduction\label{section:intro}}
Understanding the mechanisms by which living (active) systems such as cells and bacteria invade and navigate through their surroundings is of pivotal importance in many physiological and pathological processes. Depending on the physical and chemical properties of their microenvironment active matter can show distinct modes of invasion, from single cell migration to groups of cells moving as a collective. The latter has been identified as the primary mode of cancer cell invasion~\cite{Clark15,Clark19}. Similarly collective migration is the prime mode of growth and invasion by bacterial biofilms~\cite{Wu15,Even17,Hartmann18}.

Among physical environmental factors, geometrical constraints have a strong influence on the dynamics of collective migration in biological matter. Various physiological processes such as bacterial filtering rely on active matter living within pores, cavities, and constrictions~\cite{Yang2010}. {\it In vivo}, cancerous cells are known to preferentially move along pre-existing tracks of least resistance, such as myelinated axons or blood vessels, when invading into healthy tissue~\cite{Gritsenko2011,Weigelin2012}. Similarly, during biofilm formation interaction between the bacterial colony and the surface it grows on is of major importance~\cite{Hall-Stoodley2004,Conrad18}. The relevance of the interaction between biological matter and confining geometries in many physiological processes has prompted extensive experiments. 
In particular, {\it in-vitro} experiments on epithelial cells~\cite{Vedula2012,Marel2014,Marel2014_2,Yang2016,Gauquelin2019,Duclos2017,Duclos2018}, bacteria~\cite{Wioland2016,Conrad2018}, and mixtures of biofilaments and molecular motors~\cite{Suzuki2017,Wu2017} have shown that confinement to channel-like geometries generically alters the flow of biological matter in a significant way. Striking examples are the crossover from chaotic flows of bacteria~\cite{Wioland2013,Wioland2016} and microtubule/motor protein mixtures~\cite{Hardouin2019,Opathalage2019} to polarised movement along the long channel axis and the emergence of shear flow in fibroblast cells~\cite{Duclos2018} confined within rectangular geometries. Moreover, theoretical and computational studies of active matter interacting with a periodic array of obstacles have shown a reduction in the effective diffusion coefficient of active particles with increasing density of obstacles~\cite{Brun18} and modification of the active dispersion within periodic arrays in the presence of applied flows~\cite{Alonso19}.

While recent research has been primarily focused on the crosstalk between the confinement and the activity of the particles in order to determine patterns of motion, a complete understanding of the patterns of collective invasion and spreading also requires a detailed investigation of growth dynamics. From the physical perspective, such interplay between activity, confinement, and growth can be accompanied by additional complexities arising from hydrodynamic interactions between growing active matter and the confinement, orientation dynamics of elongated active particles, and emergent collective phenomena such as active turbulence~\cite{Wensink2012,Thampi2014,Giomi2015,James2018,Bratanov2015} and active topological defects~\cite{Elgeti2011,Giomi2013,Giomi2014,Saw2017}.
Due to this interconnection of complex physical processes, the mechanistic understanding of active matter invasion within geometrical constraints remains largely unexplored. Recent numerical analyses of a self-propelled particle model~\cite{Tarle2017}, neglecting orientational dynamics and hydrodynamic effects, have shown that the physics of growing active matter can explain some of the observed experimental phenomena~\cite{Vedula2012}, such as caterpillar motion of the advancing front and the enhancement of collective migration speed in thin capillaries, by assuming a coupling between the curvature of the front and the motility of the particles at the leading edge.

Here we use a generic model of active matter that accounts for hydrodynamics, orientational effects, and growth to investigate collective patterns of active matter invasion of capillaries.
We identify three different regimes of invasion, each with distinct interface shapes, flow patterns, orientational ordering, and topological defect dynamics. In particular, we show that above a certain threshold in the strength of the active forces, highly dynamic deformations are formed at the interface between the invading active matter and the surrounding medium. At higher activities, we also find a second threshold beyond which blobs of active matter begin to detach from the growing active column to enhance the invasion of free space. We explain the first crossover in terms of intrinsic hydrodynamic instabilities of the bulk active matter accompanied by additional instabilities arising from the presence of the growing interface. The latter can be understood by the ability of active stresses to overcome the stabilising effect of the surface tension and pinching off the growing interface. Together, our study reveals several possible invasion patterns and collective dynamics when active matter invades a capillary constriction thus providing a framework for further experimental investigations.

\begin{figure}[t]
 \centering
 \includegraphics[]{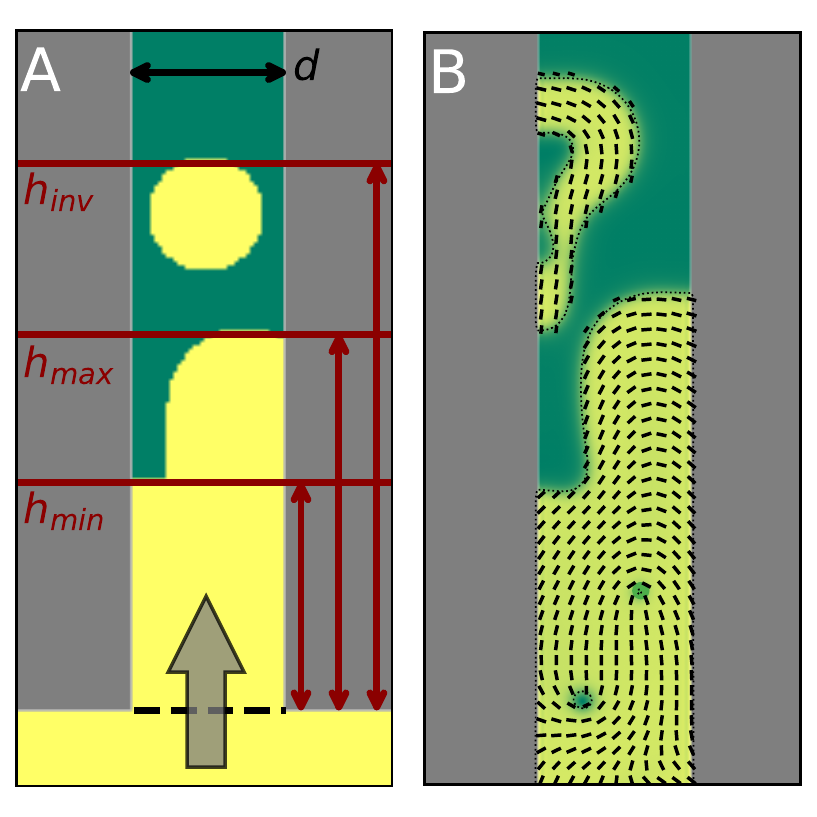}
 \caption{\emph{Simulation setup} (A) Schematic drawing of simulation setup. The active phase (yellow) invades from the broader reservoir (region below black dashed line) where it can grow into the narrower capillary (width $d$) that is initially filled with isotropic liquid (green). The broad gray arrow illustrates the invasion direction. Different heights are marked for later discussion of observables: $h_{\text{max}}$, the highest point of the active phase which is connected to the reservoir, $h_{\text{inv}}$, the highest point that any patch of active phase reaches, and $h_{\text{min}}$, the highest point where the capillary is filled with active fluid to the full width.
(B) A representative simulation snapshot to illustrate the dynamics. Black bars mark the director field; the two green speckles inside the nematic phase are cores of +1/2 topological defects.}
 \label{fig:setup}
\end{figure}

\section{Model\label{section:model}}
We employ a two-phase model of active nematohydrodynamics~\cite{Blow2014} to explore the growth of an active layer into otherwise isotropic surroundings within a confined channel and use a hybrid lattice Boltzmann (LB) method to numerically solve the equations~\cite{Marenduzzo2007}. This choice was motivated by the generality of this class of continuum models which is due to the fact that only local conservation laws, a nematic interaction between the systems' constituents, and perpetual injection of energy at the smallest length-scale are assumed. Models of this kind reproduce a variety of non-equilibrium flows such as stable arrays of vortices or chaotic flows, together with diverse stationary and non-stationary patterns of nematic ordering and topological defects~\cite{Simha2002,Thampi2013,Giomi2013,Giomi2015,Thampi2014,DeCamp2015,Cortese2018, Doostmohammadi2018}. On a phenomenological level, these are successful in modelling the collective dynamics in biological systems in cases such as microtubule/motor-protein mixtures~\cite{Sanchez2012,Giomi2013,Thampi2013}, cellular monolayers~\cite{Bittig2008,Duclos2014,Doostmohammadi2015,Saw2017}, or bacteria~\cite{Volfson2008}. 
In addition to the bulk dynamics, the interaction with walls or obstacles of different shape can add further complexity. Theoretically, using the active nematohydrodynamics framework, a transition to spontaneous flows in channels has been predicted~\cite{Voituriez2005,Edwards2009}, as well as more complex states with intricate interplay of defects and vortices, and a transition to active turbulence~\cite{Shendruk2017,Doostmohammadi2017,Norton2018}.

\subsection{Governing equations\label{section:equations}}
We consider a two-dimensional model of a two-phase system consisting of an isotropic fluid phase and an active nematic phase~\cite{Blow2014}.
 The orientational order in the nematic phase is characterised by the symmetric and traceless nematic tensor $Q_{\alpha\beta}=S\left(2n_{\alpha}n_\beta-\delta_{\alpha\beta}\right)$~\cite{deGennes} with $S$ (magnitude of the nematic order) and $n_\alpha$ (director) indicating the magnitude and the direction of the nematic order, respectively.
The relative density of the nematic phase is measured by a scalar phase field $\phi$ which is 0 in the purely isotropic and 1 in the purely nematic phase.

The free energy density of the system is given by (using the Einstein summation convention):
\begin{multline}\label{eq:freeenergydensity}
f=\frac12\binary\phi^2\left(1-\phi\right)^2+\frac12\coupling\left(\phi-\frac12Q_{\alpha\beta}Q_{\alpha\beta}\right)^2\\+\frac12\Kphi\partial_\gamma\phi\partial_\gamma\phi+\frac12\Knem\partial_\gamma Q_{\alpha\beta}\partial_\gamma Q_{\alpha\beta},
\end{multline}
where $\binary$, $\coupling$, $\Kphi$, and $\Knem$ are positive material constants. Since $Q_{\alpha\beta}Q_{\alpha\beta}=2S^2$, the second term ensures a tight coupling between the magnitude of the nematic order $S$, and the phase field $\phi$.. Together with the first term, which corresponds to a Cahn-Hilliard free energy~\cite{ChaikinLubensky,Orlandini1995}, this leads to well-defined interfaces betweens a nematic ($S=1,\phi=1$) and an isotropic ($S=0,\phi=0$) phase (Fig.~\ref{fig:setup}).
The third and fourth terms are elastic energies. The first, third, and fourth terms contribute to the surface energy and 
the fourth term also penalises bulk deformations in $Q_{\alpha\beta}$. The free energy then reads
\begin{align}
\mathcal{F}=\int\text{d}^2\vec{r}\;f.
\end{align}

The order parameters evolve according to the following equations:
\begin{align}
\partial_t\phi+\partial_\beta\left(\phi u_\beta\right)=&\Gamma_\phi\Delta\mu,\label{eq:phi}\\
\left(\partial_t+u_\kappa\partial_\kappa\right)Q_{\alpha\beta}=&-\tumbling\Sigma_{\alpha\beta\kappa\lambda}E_{\kappa\lambda}-T_{\alpha\beta\kappa\lambda}\Omega_{\kappa\lambda}+\Gamma_Q H_{\alpha\beta},\label{eq:Q}
\end{align}
with
\begin{align}
\mu=&\frac{\partial f}{\partial\phi}-\partial_\gamma\frac{\partial f}{\partial\left(\partial_\gamma\phi\right)}\\
H_{\alpha\beta}=&\left(\delta_{\alpha\beta}\delta_{\kappa\lambda}-\delta_{\alpha\kappa}\delta_{\beta\lambda}-\delta_{\alpha\lambda}\delta_{\beta\kappa}\right)\nonumber\\&
\left\lbrace
\frac{\partial f}{\partial Q_{\kappa\lambda}}-
\partial_\gamma\left(\frac{\partial f}{\partial\left(\partial_\gamma Q_{\kappa\lambda}\right)}\right)
\right\rbrace,\\
E_{\alpha\beta}=&\frac12\left(\partial_\beta u_\alpha + \partial_\alpha u_\beta \right),\:\:\:\:\:
\Omega_{\alpha\beta}=\frac12\left(\partial_\beta u_\alpha-\partial_\alpha u_\beta\right),\\
\Sigma_{\alpha\beta\kappa\lambda}=&Q_{\alpha\beta}Q_{\kappa\lambda}-\delta_{\alpha\kappa}\left(Q_{\lambda\beta}+\delta_{\lambda\beta}\right)\nonumber\\
&-\left(Q_{\alpha\lambda}+\delta_{\alpha\lambda}\right)\delta_{\kappa\beta}+\delta_{\alpha\beta}\left(Q_{\kappa\lambda}+\delta_{\kappa\lambda}\right),\\
T_{\alpha\beta\kappa\lambda}=&Q_{\alpha\kappa}\delta_{\beta\lambda}-\delta_{\alpha\kappa}Q_{\beta\lambda}.
\end{align}
The l.h.s. of Eqs.~\eqref{eq:phi} and~\eqref{eq:Q} are convective derivatives with the underlying velocity field $u_\alpha$. Advection and diffusion drive the dynamics in $\phi$; $\Gamma_\phi$ is the corresponding diffusion constant. $\Gamma_Q$ is a rotational diffusion constant which, together with the molecular field $H_{\alpha\beta}$, controls diffusive relaxation in $Q_{\alpha\beta}$. In addition, the interplay of flow $u_\alpha$ and order $Q_{\alpha\beta}$ is less trivial. The first and second terms on the r.h.s. of the Eq.~\eqref{eq:Q} form the co-rotational derivative which accounts for the response of the orientation field to the extensional ($E_{\alpha\beta}$) and rotational ($\Omega_{\alpha\beta}$) components of the velocity gradients respectively. $\tumbling$ is the tumbling parameter which determines the relative influence of the rate of strain on the director orientation. It depends on the geometry of the active particles, for prolate ellipsoids $\tumbling>0$, while for oblate ellipsoids $\tumbling<0$, and for spherical particles $\tumbling=0$~\cite{Larson99}.

The velocity field $u_\alpha$ obeys the Navier-Stokes equations:
\begin{align}
\partial_t\rho+\partial_{\beta}\left(\rho u_{\beta}\right)=&0,\\
\rho\left(\partial_t+u_{\beta}\partial_{\beta}\right)u_{\alpha}=&\partial_{\beta}\Pi_{\alpha\beta},
\end{align}
where $\rho$ is the density of the fluid and $\Pi_{\alpha\beta}$ is the stress tensor comprising viscous stress, pressure contribution, elastic stresses, and the active stress:
\begin{align}
\Pi^{visc}_{\alpha\beta}=&2\rho\eta E_{\alpha\beta},\\
\Pi^{p}_{\alpha\beta}=&-\frac\rho3\delta_{\alpha\beta},\\
\Pi^{el,1}_{\alpha\beta}=&\left(f-\mu\phi\right)\delta_{\alpha\beta}\nonumber\\&-\frac{\partial f}{\partial\left(\partial_\beta\phi\right)}\partial_\alpha\phi-\frac{\partial f}{\partial\left(\partial_\beta Q_{\kappa\lambda}\right)}\partial_\alpha Q_{\kappa\lambda},\\
\Pi^{el,2}_{\alpha\beta}=&\left(\tumbling\Sigma_{\alpha\beta\kappa\lambda}+T_{\alpha\beta\kappa\lambda}\right)H_{\kappa\lambda},\\
\Pi^{act}_{\alpha\beta}=&-\zeta \phi Q_{\alpha\beta}.
\end{align}
Here, $\eta$ is the viscosity, and the elastic stresses $\Pi^{el,i}_{\alpha\beta}$ describe feedback from variations in the order parameters on the fluid flow~\cite{Blow2014}. The definition of the active stress $\Pi^{act}_{\alpha\beta}$ is such that any gradient in $Q_{\alpha\beta}$ generates a flow field and drives the system at small length-scales, with strength determined by the magnitude of the activity $\zeta$. A positive (negative) $\zeta$ corresponds to an extensile (contractile) material. The dipole flow-fields generated with this ansatz correspond to those of microswimmers - ``pushers'' generate extensile stresses, ``pullers'' contractile stresses~\cite{Ramaswamy10}. The active stress continuously drives the system out of thermodynamic equilibrium.
It will be of major importance in the following discussions.

\begin{figure*}[ht]
 \centering
 \includegraphics[]{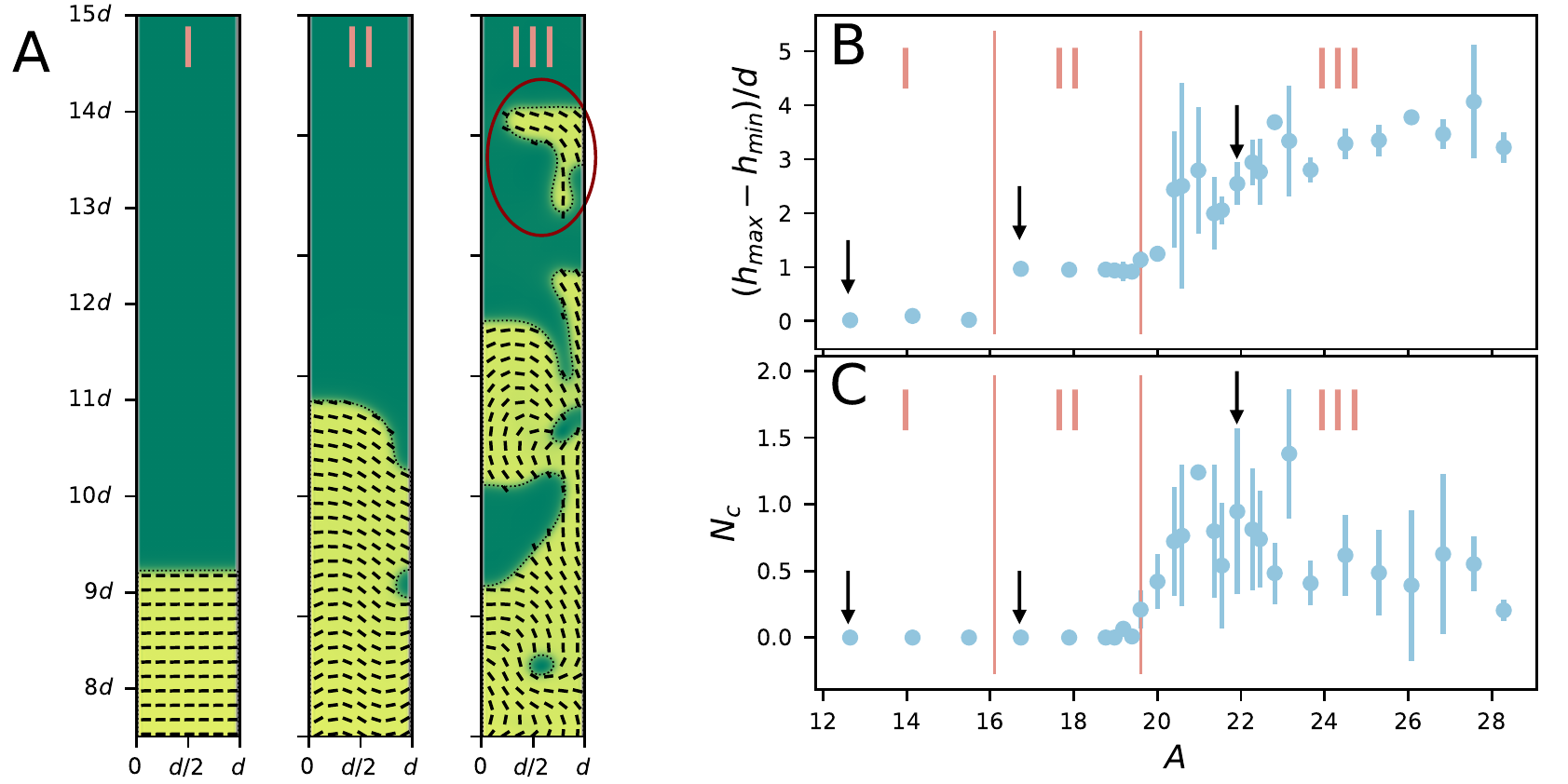}
 \caption{\emph{Phenomenological regimes for varying activity number.} (A) Snapshots from simulations showing typical configurations of the different regimes. From left to right: Regime I with flat interface, regime II with a deformed interface, and regime III where clusters (red circle) start to detach from the main active phase. Corresponding ESI Movies are \ESImovieI, \ESImovieIII, and \ESImovieVI, respectively, see ESI section~\ESIsectionmovies\dag. (B) Quantitatively, the crossover from regime I to II is characterised by a sudden increase in $\hdiff$ from $0$ to $1$ (see Fig.~1 for the definition of $h_{\text{max}}$ and $h_{\text{min}}$). By contrast, within regimes I and II, this quantity is nearly constant. In regime III it increases with activity. (C) While there is only one single coherent patch of active material in regimes I and II, the defining property of the crossover from II to III is the appearance of additional clusters. $N_c$ denotes the number of clusters in addition to the bulk active phase. Black arrows in B and C mark values for which snapshots in A were taken. Errorbars depict 1.96 SEM (standard error of the mean). To estimate the SEM, simulations with different initial noise were performed for identical parameter configurations.}
 \label{fig:regimes}
\end{figure*}

\subsection{Simulation setup\label{section:setup}}
In Fig.~\ref{fig:setup}, we show the geometry for which we will solve the equations. This simulation setup corresponds to the experimental configurations for cell monolayers in references~\cite{Vedula2012,Marel2014,Marel2014_2,Yang2016,Gauquelin2019}. It consists of a wider reservoir that feeds a narrower capillary. At $t=0$, the active phase is restricted to the large reservoir.

In order to generate new active material in the reservoir, we locally increase $\phi$ to values higher than $1.0$ at random positions. For every point in the active phase in the reservoir a \textit{growth event} takes place with probability $r\phi(1 - \phi/\phi_{c})$.  This means that for a given time $\tau_{g}$, a source term $\alpha\phi$ is added to the r.h.s. of Eq.~\eqref{eq:phi} in a circle of radius $r_{g}$ around this site. Diffusion and convection then lead to a spreading of the active phase, causing it to rise into the thinner capillary.
This local implementation leads to additional flow in the reservoir as growth events generate dipole-like flow fields and a growth pressure through the isotropic part of $\Pi^{el,1}$. The logistic-growth-like saturation in the probability $r\phi(1 - \phi/\phi_{c})$ prevents unbounded growth of $\phi$. We find that the details of this implementation or even the geometry of a reservoir are not important for the qualitative dynamics in the capillary (see ESI section~\ESIsectionreservoir\dag). 

In this study, we use the following parameters: $\binary=0.08$, $\coupling=0.15$, $\Knem=0.02$, $\Kphi=0.08$, $\Gamma_\phi=0.2$, $\Gamma_Q=0.4$, $\tumbling=0.7$, $\eta=1/6$, $r=0.001$, $\tau_{g}=10000$, $r_{g}=5$, $\alpha=0.01$, $\phi_{c}=1.2$, unless otherwise noted. We vary $\zeta$ from $0.0020$ to $0.0100$ to investigate how invasion depends on the strength of the active driving. The free energy coefficients are chosen such that there is a well-defined interface between the active and the passive phase. $\tumbling=0.7$ leads to alignment of the director to an external flow in a passive nematic. 
Parameter fitting of the continuum equations to physical active systems remains a topic of research; therefore, we consider a generic parameter set that has been shown to reproduce the flow vortex-lattice generated by a dense assembly of endothelial cells~\cite{Rossen2014,Doostmohammadi2016NC}, and the flow fields of dividing Madin-Darby Canine Kidney cells\cite{Doostmohammadi2015}.
Independent samples for parameter sets are obtained by varying the initial condition randomly.
Boundaries are no-slip with respect to the fluid and von-Neumann for $\phi$ and $Q_{\alpha\beta}$, meaning that there is no preferred anchoring of the director in the absence of activity. Variation of the boundary conditions can potentially lead to the emergence of additional exotic phases of invasion, but their systematic study lies beyond the scope of this work. For the reservoir, we have periodic boundary conditions in the direction perpendicular to the capillary axis.

In the following, we characterise the strength of the activity by the dimensionless activity number $A = d/(\sqrt{K_{\text{Q}}/\zeta}) = d/\Lambda_{\zeta}$.
The activity-induced length-scale $\activescale=\sqrt{\Knem/\zeta}$ emerges from the competition between the activity driving the dynamics and the elastic resistance against deformations in the director field~\cite{Thampi2014,Hemingway2016,Guillamat17}, while the capillary width $d$ imposes an upper limit for hydrodynamic interactions across the capillary.
The activity number $A$ relates these two length-scales~\cite{Shendruk2017,Doostmohammadi2017}.

\section{Results\label{section:results}}

\subsection{Distinct invasion regimes}
For a first characterisation of the invasion behaviour of active matter into a capillary, we will focus on how the phenomenological changes in the structure and the dynamics of the interface between the active phase and the isotropic fluid depend on the dimensionless activity $A$. Surprisingly, we observe that this can be categorised into three different regimes separated by two well-defined crossovers under variations of this single quantity alone. 

Figure~\ref{fig:regimes}A shows characteristic snapshots for the three different regimes. For small activities, below a certain threshold ($A\sim16$), the system is characterised by a flat interface between the isotropic and active phase that advances steadily from the reservoir. We denote this as regime I. Due to the extensile activity of the particles (i.e. $\zeta>0$), the active stresses generate a preferential orientation of the director parallel to the interface, an effect called `active anchoring'~\cite{Blow2014}. For low $A$, the director field remains homogeneous throughout the nematic phase and the hydrodynamic instabilities are suppressed. 

When approaching the crossover from regime I to regime II by increasing the activity, bend deformations of the director field arise in the bulk without influencing the shape of the interface  (see ESI Movies~\ESImovieI ~and \ESImovieII\dag).
The defining property of regime II is the crossover to a state where the interface is deformed to an S-shape while keeping a $90^{\circ}$ contact angles at both walls.
As discussed in more detail in the next section, within this regime the deformed interface initially advances with the leading point on one side of the capillary wall (see ESI Movie~\ESImovieIII\dag). With a further increase in activity, we observe a periodic switching of the S-shape of the interface from one side of the capillary to the other side as the system approaches the third invasion regime.
 
In both regimes, I and II, the active phase remains as one single coherent phase which is connected to the reservoir. This changes with the crossover to regime III at $A\sim20$, where a higher activity leads to the dispatching of small clusters of active material from the main body of the active phase, that protrude deeper into the capillary. In addition, the active interface and the director field are strongly deformed and highly dynamic. Defects are regularly created at the boundary and move throughout the bulk of the active phase (see ESI Movie~\ESImovieVI\dag).

\begin{figure}[t]
\centering
 \includegraphics[]{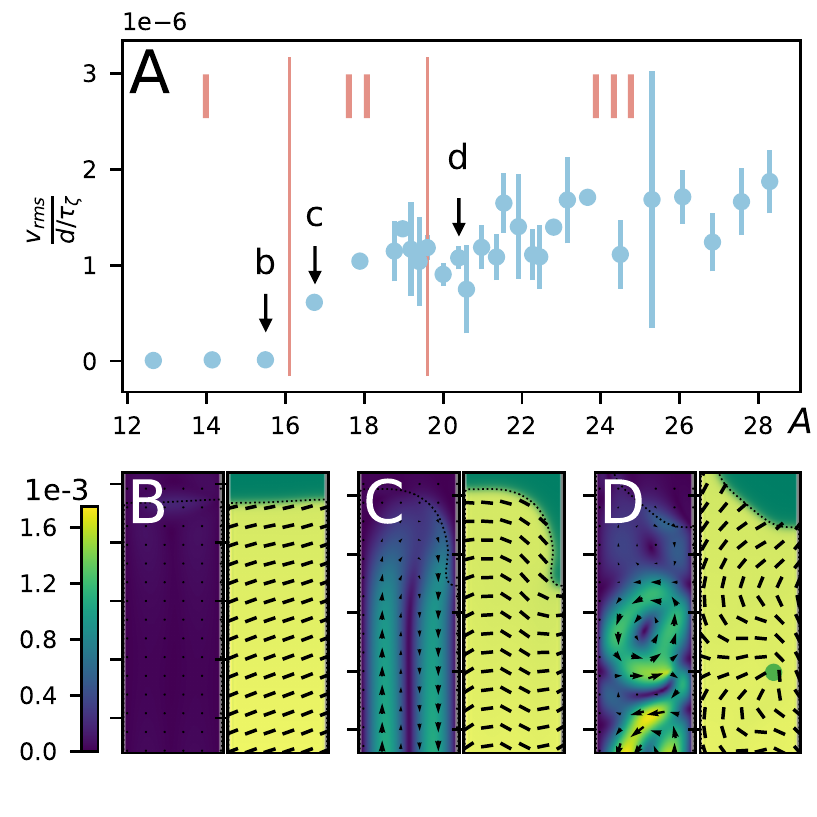}
 \caption{\emph{Comparison of flows across regimes.} (A) Root mean-squared velocity in units of $d$ over the active time scale $\tau_\zeta=\eta/\zeta$~\cite{Orlandini1995} plotted against $A$ . Crossover from regime I to II coincides with the appearance of finite flows. Strength of flows increases inside regimes II and III with a small plateau at the crossover from regime II to regime III. Lower case letters indicate the values at which examples in (B-D) were taken. (B-D) Characteristic velocity and orientation fields near the interface. (B) For $A=15.5$, the system is in regime I, the interface is flat with no flows near it. (C) At $A=16.7$, in regime II, the interface is bent and the characteristic flow that turns at the interface is visible. (D) In regime III ($A=20.4$), the interface can take various forms, $+1/2$-defects (green dot) appear in the bulk and the flow becomes highly dynamical and complex. ESI Movies corresponding to panels B-D are \ESImovieII, \ESImovieIII, and \ESImovieV, respectively, see ESI section~\ESIsectionmovies\dag.}
 \label{fig:flows}
\end{figure}

\subsection{Detailed characterisation of  invasion regimes and crossovers\label{section:results_quantitative}}
To understand the connection between the different interface dynamics and the overall invasion process, we further characterise the defining features of the three different regimes. Quantitatively, the crossovers between the regimes can be read off from two observables: (i) $\hdiff$, which characterises deformations of the isotropic-active interface in the capillary (see Fig.~\ref{fig:setup}A for definition of $h_{\text{max}}$ and  $h_{\text{min}}$), and (ii) $N_c$ which is the number of clusters of the active phase that are in the system in addition to the main body of the active phase connected to the reservoir.
As the activity number is increased, the crossover from regime I to regime II is reflected by a jump in the mean value of $h_{\text{max}}-h_{\text{min}}$ from 0 to the width $d$ of the capillary, or equivalently a jump of $\hdiff$ from 0 to 1 (Fig.~\ref{fig:regimes}B), which mirrors the S-shape of the interface in regime II. Within the parameter range studied here, we did not observe any variation in the positioning of the crossover from regime I to regime II under variation of $\binary$. The second crossover, from regime II to regime III, is also marked by a further jump in $\hdiff$, but is even more evident in $N_c$ which becomes finite (Fig.~\ref{fig:regimes}C), reflecting the emergence of additional smaller clusters that detach from the main active phase. It is noteworthy that a similar quantitative behaviour is observed when varying the width of the channel (see ESI Fig.~\ESIfigwidth~in ESI section~\ESIsectionwidth\dag), indicating that the activity number is the relevant dimensionless parameter in this setup.

\subsubsection{Properties of regime II.~~} 
The crossover from regime I to II is also accompanied by sharp changes in the flow and director fields in the capillary. This not only sheds light on the underlying mechanism, but also has consequences for the overall speed of invasion as we will see later (section \ref{section:invasion}). Figures~\ref{fig:flows}A,B show that below the first crossover, there are no flows in the capillary ($v_{\text{rms}}=0$). For higher activities, with a non-zero $\hdiff$, finite flows are generated (Figs.~\ref{fig:flows}A,C). This is reminiscent of the well-established spontaneous flow transitions in confined active matter~\cite{Voituriez2005,Marenduzzo2007}, which arise due to the formation of hydrodynamic instabilities in active nematics above a certain threshold of activity. While in those cases the active force alone is responsible for the transition, in the present situation the growth dynamics in the reservoir induces long-range effects which influence the position of the crossover (for a system without reservoir, see ESI section~\ESIsectionreservoir\dag). The advection of active material along with these flows causes the characteristic S-shape of the interface in this regime. Interestingly, $v_{\text{rms}}$ is increasing with higher activity numbers within regime II while $\hdiff\sim1$ throughout.

\begin{figure}[t!]
\centering
 \includegraphics[]{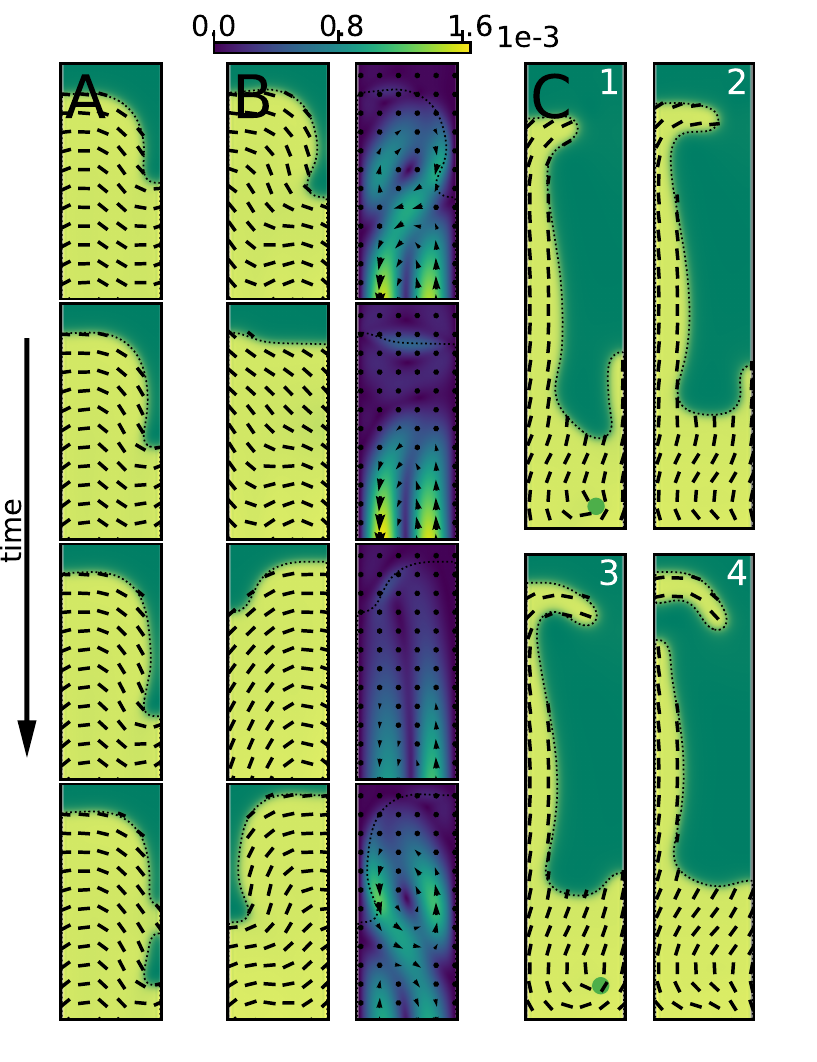}
 \caption{\emph{Interface dynamics in regimes II and III.} 
(A) Characteristic periodic motion in regime II ($A=17.9$). The $+1/2$-defect like bend-deformation on the right wall moves backwards relative to the interface until it is pinched off and dissolved in the nematic bulk. Simultaneously the interface relaxes back to a less stretched state and the cycle begins anew. 
(B) Switching in the S-shape for higher activities in regime II ($A=19.2$). When approaching the second crossover, vortices emerge in the capillary (panels on the right). When the most forward vortex changes its rotation direction, the director follows and bend deformations as well as interface shape adapt accordingly. 
(C) Time series of a cluster-formation event in regime III ($A=20.4$). The nematic layer at the wall becomes unstable until finally a cluster detaches from the main active phase. 
In each of (A-C) the simulation time between all sequential snapshots is equal and is measured in units of the active time scale $\tau_\zeta=\eta/\zeta$~\cite{Thampi2015}, A: $\delta t=480\tau_{\zeta}$, B: $\delta t=2760\tau_{\zeta}$, C: $\delta t=1248\tau_{\zeta}$. ESI Movies corresponding to panels B and C are \ESImovieIV~and \ESImovieV, respectively, see ESI section~\ESIsectionmovies\dag.}
\label{fig:dynamics}
\end{figure}

Moreover, because of the impact of the extensile activity that keeps the director aligned with the interface, the increasing bend deformation looks similar to a backwards pointing $+1/2$-topological defect at the less advanced side of the interface (Fig.~\ref{fig:flows}C; right panel). Corresponding to the characteristic self-motility of $+1/2$-topological defects~\cite{Putzig2014,Shendruk2017,Saw2017}, the interface on this side continues to move backwards relative to the mean interface-position stretching out the interface until the defect detaches from the interface and moves into the active phase. Concomitantly with this defect absorption into the nematic, the interface re-attaches to the wall at a higher position. Repeatedly, this leads to a periodic shape-change in the interface (see Fig.~\ref{fig:dynamics}A and the ESI Movie~\ESImovieIII\dag).

As the activity is increased further, the initial spontaneous transition to flows is followed by the generation of counter-rotating vortices along the capillary. As the active phase grows and advances through the capillary, a lattice of vortices is formed behind the interface, reminiscent of vortex-lattices in a non-growing confined active matter~\cite{Doostmohammadi2017,Theillard17}, which appear when the activity-induced vorticity length scale becomes comparable to the capillary width. Interestingly, however, in this growing active matter, the orientation of the vortex closest to the interface also determines the parity of the interface. As a consequence, a change in the orientation of the most forward vortex leads to a switching at the interface, with the S-shape of the progressive front flipping from one side of the channel to the other with respect to the capillary axis (see Fig.~\ref{fig:dynamics}B and the ESI Movie~\ESImovieIV\dag).

Within regime II, the rms-velocity $v_{\text{rms}}$ grows approximately linearly with the activity number $A$ and the flow field remains structured until transitioning to active turbulence in regime III (Fig.~\ref{fig:flows}D).

\subsubsection{Properties of regime III.~~} 
While the crossover to regime II was solely controlled by the hydrodynamic instability to spontaneous flow formation within the bulk of the active matter in the capillary, the crossover from regime II to III, where active clusters appear, is strongly influenced by the competition between activity and surface tension at the interface.
This interfacial effect can be clearly seen in a close-up view of the interface at the onset of cluster detachment (Fig.~\ref{fig:dynamics}C): At a sufficiently high activity number relatively long and thin layers of active matter with an approximately uniform director field are formed at the sides of the capillary wall. As the simulation time goes by, a small deformation at the tip of the layer develops and grows until it breaks away from the active layer, forming a detached cluster  (see ESI Movie~\ESImovieV\dag). The formation of clusters means an increase in total interface length which is increasingly unfavourable for higher surface tension and thus requires higher active stresses to be generated. Therefore, the breakup of the thin layer can be understood as the destabilising effect of the activity that leads to the interface deformation dominating over the stabilising effect of the surface tension, working to keep the interface straight. However, especially in this complex setup, it is not possible to decide to what degree this effect is purely interfacial and whether dynamics in the bulk are important. The simultaneous appearance of $+1/2$ topological defects in the nematic phase (see ESI Fig.~\ESIfigdefects\dag) hints that changes in the bulk dynamics are also involved in the crossover as flow dynamics become more reminiscent of turbulent dynamics and higher active stresses are generated. Interestingly, directly after the crossover the number of additional clusters peaks. With a further increase in $A$ the clusters are still present, however the number of clusters goes down. A close look at the interface reveals that at very high activities the active material forms extremely long and thin layers on the capillary walls and the detached clusters quickly reattach to the main body of the active nematic, leading to a smaller number of clusters on average.

\subsection{Invasion capability\label{section:invasion}}
\begin{figure}[t]
\centering
 \includegraphics[]{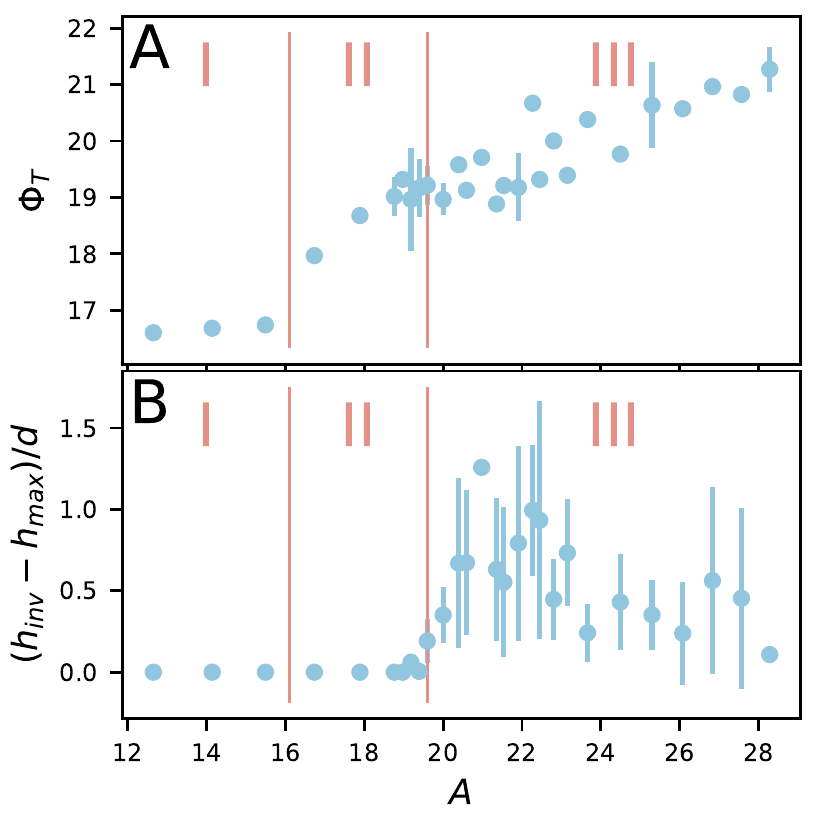}
 \caption{\emph{Propensity of the system to invade the capillary.} (A) Invasion index $\Invasion:=\int \phi\left(\mathbf{r},T\right)\:\text{d}^2\mathbf{r}/d^2$ defined as the amount of active material inside the capillary after $9.5\:10^6$ simulation steps plotted versus activity number $A$. In regime I, the invasion shows little dependence on $A$ until the appearance of flows at the crossover from regime I to II leads to a sharp increase. (B) Relative difference of the maximum invasion height and the maximum height of the interface $\hdiffinva$ against $A$. In regimes I and II, this difference is trivially zero. In regime III it shows an increase of up to one capillary-width meaning the clusters protrude significantly deeper into the capillary than the main body of the growing active matter.}
 \label{fig:invasion}
\end{figure}

In the context of our original research question we now investigate the influence of the different regimes on the capability of the active system to claim new territories. To this end, we show in Fig.~\ref{fig:invasion}A the total amount of active material in the capillary at the end of the simulation ($9.5\:10^6$ simulation steps) $\Invasion:=\int \phi\left(\mathbf{r},T\right)\:\text{d}^2\mathbf{r}/d^2$, which we denote as the invasion index. In the absence of flows (regime I, when the process is purely diffusive), the invasion index $\Phi_T$ is almost independent of $A$ and it is significantly lower than in regimes II and III, where activity-induced flows enhance the advective transport of active material from the reservoir to the capillary. After the crossover from regime I to II, $\Invasion$ follows a linear dependence on $A$ similar to the linear increase of $v_{rms}$ in this region.
The crossover from regime II to III does not alter this quantity in a qualitative way, i.e. the additional clusters have no influence on the total amount of active material that invades the capillary. However, they can be shown to invade about $0.5$ to $1.0$ capillary widths deeper into the capillary than the tip of the coherent active phase connected to the reservoir. This is best illustrated by plotting the relative difference of the maximum invasion height $h_{\text{inv}}$ (see Fig.~\ref{fig:setup}A) and the maximum height of the interface $h_{\text{max}}$: $\hdiffinva$ (Fig.~\ref{fig:invasion}B).
Together, these results indicate that the flows in the capillary and the interfacial dynamics control the rate at which active material enters the capillary and thus determine the invasion speed.

\section{Conclusions\label{section:conclusion}}
In this work, we have presented a two-phase computational framework for studying the growth of active matter within an isotropic fluid medium. Our approach accounts for hydrodynamic effects and the orientational dynamics of active particles, and is able to reproduce emergent active phenomena including spontaneous flow generation and active turbulence. Motivated by the prevalence of active matter growing within confined spaces and constrictions, we investigated the dynamics in capillaries of different sizes, focusing on the combined effects of growth and the active stresses that are continuously exerted by active particles on their surroundings.

We find three distinct invasion patterns as the activity is varied: (i) at small activities invasion is completely controlled by the growth dynamics and the interface between the active phase and the surroundings remains stable with no, or slight, deformations. (ii) Increasing activity beyond a given threshold results in spontaneous flow generation which significantly enhances the invasion within the capillary and is accompanied by deformations of the interface. (iii) At yet higher strengths of activity, a second crossover appears, where active clusters begin to detach from the main body of the invading active phase, further enhancing the invasion of the surrounding space.

Our analysis shows that the crossovers between the various invasion regimes are controlled by different mechanisms. The crossover between regimes I and II is governed by the well-established spontaneous flow formation in confined active matter~\cite{Voituriez2005,Edwards2009}, which is due to a hydrodynamic instability of the active nematic. We show that the onset of flow within the bulk is accompanied by the deformation of the interface and can even result in a periodic switching of the active protrusions from one side of the capillary to the other, as flow vortices start to form behind the advancing front.
When activity becomes stronger, turbulent flows and defects are observed in the bulk of the active material and, dependent on its surface tension, the system also becomes able to rip apart the active-to-passive interface marking the third invasion regime.

Our results thus highlight the significant differences in invasion patterns when the activity of growing matter is accounted for. From a biological perspective, this could be linked to the invasive behaviour of biological matter such as growing colonies of cells or bacteria. It suggests that qualitative changes in the spread of these organisms into confined spaces can be caused by changes in the availability and conversion rate of chemical energy into mechanical stresses. One could conjecture that more invasive cell lines or bacterial strands could regulate their invasion of pores or cavities by tuning their strength of motion. For example, inducing the formation of additional clusters might be beneficial for cell-lines that aim at aggressive invasion of surrounding tissue while, for bacteria, it might be beneficial to remain coherent in order to profit from the benefits of cooperative behaviour.

Considering possible experimental realisations, our model predictions apply to active growing systems, where hydrodynamic interactions and nematic orientational order play an important role. Such nematic ordering has been reported in various biological systems including bacterial colonies~\cite{Volfson2008,Li19}, cultures of amoeboid cells~\cite{Gruler99}, fibroblast cells~\cite{Duclos2014,Duclos2017}, human bronchial cells~\cite{Blanch18}, neural progenitor stem cells~\cite{Kawaguchi2017}, and Madin-Darby Kanine Kidney (MDCK) epithelial monolayers~\cite{Saw2017,Saw18}. Investigating the degree to which experiment and theory can be matched by the adaptation of parameters and boundary conditions in the model or the reasons for deviations is a powerful way of unraveling the role of mechanics in determining the behaviour of such active biological matter. It would also be interesting to compare to models that resolve individual particles such as models of self-propelled particles~\cite{Tarle2017}, phase field approaches~\cite{Shao2010,AransonBook,Mueller2019}, or cellular Potts models~\cite{Thueroff2019}. In contrast to these, our model by construction cannot resolve phenomena or implement mechanisms that act on the single particle-level, for example dynamics of cell size or shape, but due to its generality could contribute to a deeper understanding of the generic fundamental mechanisms that underlie the invasion of microscopic biological systems. 

Finally, several improvements can be envisaged for the current model. In the context of growing tissues, studies have highlighted the importance of leader cells at the progression front~\cite{Poujade07,Trepat09}. Within particle-based models such an effect is shown to be captured by introducing a curvature-dependent motility for the cells at the interface~\cite{Mark10,Tarle2017}. Within our proposed framework, a similar effect could be modelled by introducing a curvature dependence to the activity coefficient.
In addition, the natural environment of active material is often more complicated than the isotropic fluid considered here and active invasion happens through viscoelastic media. For example, cells invade the interconnected networks of collagen matrices and bacteria secrete their own extracellular matrices. The continuum framework presented here could be extended to include viscoelasticity by introducing additional order parameters representing polymer conformation.

\footnotesize
\section*{Conflicts of interest}
There are no conflicts to declare.

\section*{Acknowledgements}
E.F. and F.K. would like to thank Sophia Schaffer, Matthias Zorn, and Joachim R\"adler for stimulating discussions. A.D, F.K., and J.M.Y. would like to thank Kristian Thijssen for helpful discussions. This research was supported by the German Excellence Initiative via the program ``Nanosystems Initiative Munich'' (NIM) and  the Deutsche Forschungsgemeinschaft (DFG) via project B03 within the Collaborative Research Center (SFB 1032) ``nanoagents''.
R.M. was supported by grant P2EZP2\_165261 of the Swiss National Science Foundation. A.D. was supported by the Royal Commission for Exhibition of 1851 Research Fellowship.

The preprint template has been taken from \url{https://github.com/brenhinkeller/preprint-template.tex/}

\normalsize
\bibliography{./all_reports.bib} 
 \cleardoublepage
\appendix

\section*{Supplementary Materials}

\section{Role of reservoir and growth\label{appendix:reservoir}}
In this section we examine the importance of the cell division and the presence of a reservoir on the dynamics reported in the main text.
ESI Figure \ref{fig:reservoir} shows simulation results for two systems that differ from the one considered in the main text, by the absence of a wider reservoir, and by the absence of cell division, respectively. We see that growth is an essential factor to create the phenomena reported in the main text as in the absence of growth the behaviour is qualitatively different. If the reservoir is absent, meaning that there is only a capillary of uniform width where growth takes place in the lower region, we in contrast observe the first crossover at the same values for the activity $A$ (ESI Fig.~\ref{fig:reservoir}A), together with the same change in the invasion speed  (ESI Fig.~\ref{fig:reservoir}B). The second crossover is however significantly shifted to higher values of $A$ (see ESI Fig.~\ref{fig:reservoir}C,D), indicating that the reservoir has long-range effects on the dynamics in the capillary. These, however, do not change the qualitative picture.

\section{Universality for different channel widths\label{appendix:width}}

As described in the main text, the activity number $A$ characterises the ratio of the channel width $d$ to the intrinsic length scale of the active phase. 
In ESI Figure~\ref{fig:universality} the observables $\hdiff$ and $N_c$ are plotted against $A$ for varying channel width $d$. Both crossovers happen at the same values for $A$, independent of $d$. In addition, both observables are of the same order of magnitude and within regime II $\hdiff\sim1$. These results indicate that the activity number is the relevant dimensionless parameter in this setup and that $\hdiff$ is a meaningful observable.

\section{Description of Movies}

All movies show the time-evolution of the velocity field on the left hand side and the orientation field on the right hand side, in the same way as panels B-D in Figure~\maintextfigflows. We choose different values of the activity $\zeta$ to display the qualitatively different phenomena. There are the following files:

Higher quality movies can be found on \href{https://doi.org/10.1039/c9sm01210a}{\color{blue}{https://doi.org/10.1039/c9sm01210a}}.

\begin{description}
\item[flows0020.mp4\label{movie:0020}] $\zeta=0.0020$. System is in regime I, director homogeneous and parallel to the interface. Corresponds to Figure~\maintextfigregimes A, leftmost panel.
\item[flows0030.mp4\label{movie:0030}] $\zeta=0.0030$. System is in regime I, but the director starts to show deviations from the perfectly homogeneous configuration, especially close to the reservoir. Corresponds to Figure~\maintextfigflows B.
\item[flows0035.mp4\label{movie:0035}] $\zeta=0.0035$. System is in regime 2, showing characteristic flow- and director-fields. At later times the periodic dynamics described in section~\maintextsectionresultsquantitativeregimeII can be observed. Corresponds to Figures~\maintextfigregimes A (middle panel) and~\maintextfigflows C.
\item[flows0046.mp4\label{movie:0046}] $\zeta=0.0046$. System is the higher activity region of regime II. The video shows two examples of the switching behaviour described in section~\maintextsectionresultsquantitativeregimeIII and Figure~\maintextfigdynamics B.
\item[flows0052.mp4\label{movie:0052}] $\zeta=0.0052$. System is in regime III. This example shows the typical birth of small active clusters, corresponding to the situations shown in Figures~\maintextfigflows D and~\maintextfigdynamics C.
\item[flows0060.mp4\label{movie:0060}] $\zeta=0.0060$. System is deep in regime III. The dynamics is more turbulent than in the video for $\zeta=0.0052$; there are more defects and clusters. Corresponds to the example shown in Figure~\maintextfigregimes A on the right panel.
\end{description}

\section{Supplementary figures}
\begin{suppfigure*}[htb]
 \centering
\includegraphics[scale=1]{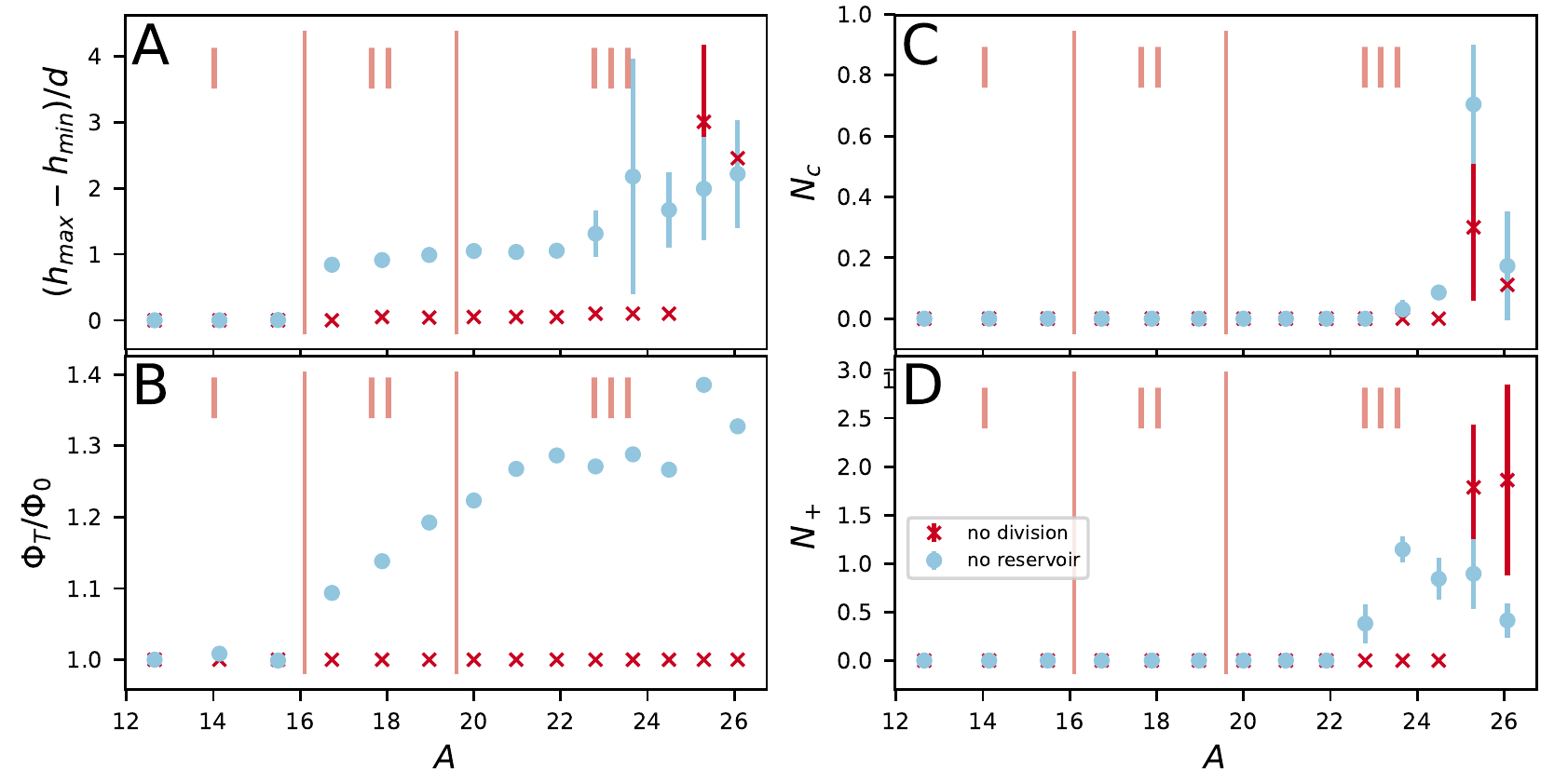}
\caption{\emph{Comparison of different modifications of the setup.}
(A) $\hdiff$ as a function of $A$. The crossover from regime I to regime II takes place at the same values as in the unmodified setup for a system without reservoir but with growth (blue circles), but is absent for a system lacking growth (red crosses). Vertical lines mark the crossover points in the unmodified setup.
(B) Total amount of active material in the system at the end of the simulation relative to the starting value.  In the non-growing system, the amount of active material is trivially constant over time. The system without reservoir shows the same qualitative dynamics as the unmodified setup. Data in this figure has been taken from one sample per value for $A$ only.
(C) The number of clusters $N_c$ as a function of $A$. The crossover from regime II to regime III takes place at higher values as in the unmodified setup for a system with reservoir but without growth, which also explains the longer linear range in subfigure B. Clusters are also present for a system lacking growth, but it takes longer for them to appear.
(D) The number of $+1/2$ defects $N_+$ as a function of $A$. The appearance of topological defects, for both setups, coincides with the appearance of additional clusters (compare subfigure C).\label{fig:reservoir}}
\end{suppfigure*}
\begin{suppfigure*}[htb]
 \centering
\includegraphics[scale=1.]{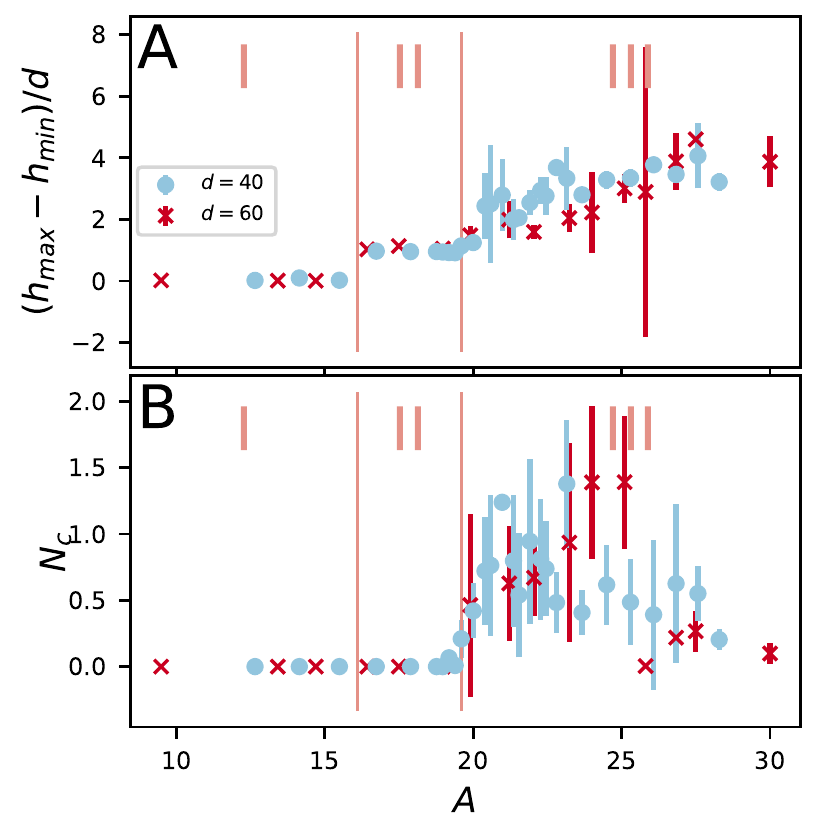}
\caption{\emph{Universality for different channel widths.} $\hdiff$ (panel A) and $N_c$ (panel B) against $A$. Both crossovers happen at the same values for both channel widths.\label{fig:universality}}
\end{suppfigure*}
\begin{suppfigure*}[htb]
\centering
 \includegraphics[]{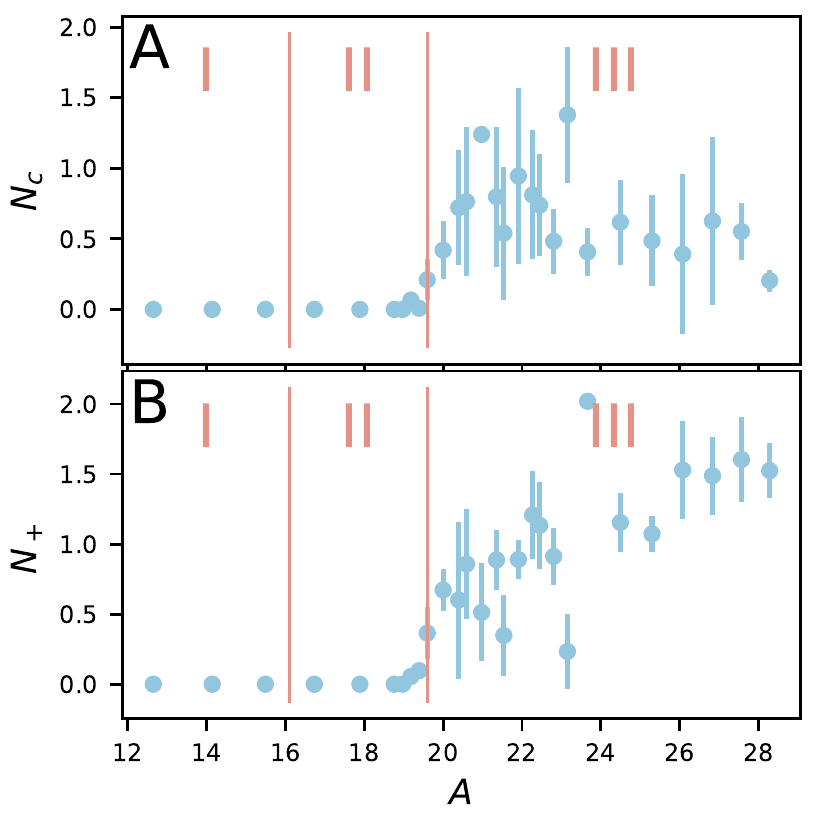}
 \caption{\emph{Transition between regimes II and III.} Comparison of number of clusters $N_c$ (A) and number of $+1/2$ topological defects $N_+$ in the system (B). Appearance of clusters and defects coincide around the second crossover.}
 \label{fig:defects}
\end{suppfigure*}
\begin{suppfigure*}[htb]
 \centering
\includegraphics[scale=1]{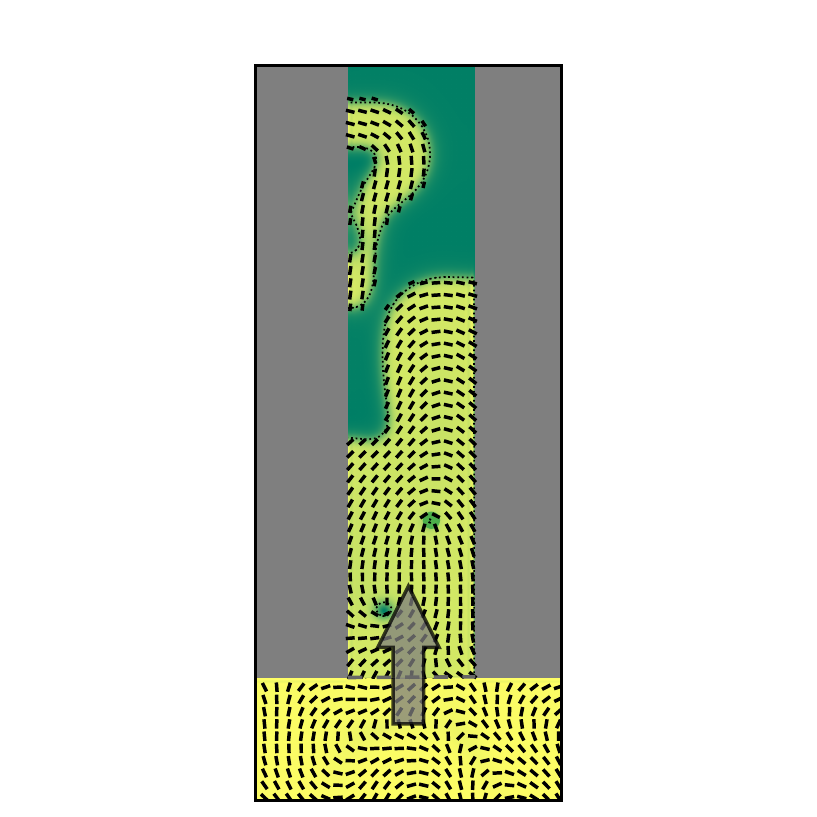}
\caption{\emph{Graphical abstract.} Biological materials such as bacterial biofilms and eukaryotic cells combine their intrinsic activity with growth dynamics to create distinct patterns of motion for invading confined spaces.}
\end{suppfigure*}

\end{document}